# Towards an unifying perspective of the fundamental properties and structural principles governing the immune system

*Paolo TIERI , Gastone C. CASTELLANI, Claudio FRANCESCHI*
"Luigi Galvani" Interdepartmental Center for Bioinformatics, Biophysics and Biocomplexity, Via San Giacomo 12, Bologna, 40126, Italy
[p.tieri; gastone.castellani; claudio.franceschi] @unibo.it

In the study of the basic properties observed in the immune system and, in a broader view, in biological systems, several concepts have already been mathematically formulated or treated in an analytical perspective, such as degeneracy, robustness, noise, and bow tie architecture [1,2,3,4,5]. These properties, among others, seem to rule many aspects of the system functioning, and share among themselvesseveral characteristics, intersecting each other, and often becoming one the indivisible part of the other. According to Kitano [2], systems biology needs solid theoretical and methodological foundation of principles and properties, able to lead towards a unified perspective. An effort in unifying the formalization and analysis of these principles can be now timely attempted.

**Degeneracy** in biological networks and neural systems is generally defined as ***the ability of elements that are structurally different to perform the same function*** (output), opposed to redundancy, when the same function is performed by identical elements [1,6]. In a more formal acception, degeneracy is a measure of the mutual dependence between subsets of elements in a system and their outputs. In other words, the measure of the degeneray can tell something about how a given configuration of the elements are qualitatively and quantitatively related with the resulting output. A degenerated system shows a certain degree of redundant functionality, maintaining at the same time the capability -due to the diversity of the elements that compose it- of yielding different outputs. In other words, many different elements can affect the output in a similar way and, together, can still have independent effects. Hence, the functional redundancy and integration of the system is coupled with the differentiation of its elements, that in turn results in the differentiation of the outputs, features that make the system more complex. On the contrary, a highly redundant system is not able to yield different outputs, given the identical nature of its elements. The nature itself, as well as the higher complexity, of a degenerated system makes it adaptable to unpredictable changes of the environment, a characteristics that has shaped in time the complex biological systems and subsystems, such as the immune system [1,6,7,8,9,10].

**Robustness** of a system is ***the capability of maintaining functions against perturbations*** [2]. It should be distinguished from stability and -more common in biology- homeostasis, since this capability can be also achieved by switching among different steady states that the system can get. More, robustness can be achieved also by means of instability, as in the case of high mutation rates due to high chromosome instability that make tumors resistant to chemotherapies [2,11,12]. Robustness against a given perturbation often implies compromises, since increase of resistance requires a resource demand that impinges both on increased fragility against other perturbations, and on degradation of performance. This tradeoff between robustness and fragility can in some way configure a sort of conservation law, inspired by the Bode law on frequency sensitivity of amplifiers [13]. Speaking in mathematical terms, robustness has been defined [2] as ***the probability that a given perturbation appears, multiplied by a function that describes the capability of the system to resist to that perturbation, integrated over the space of all the possible perturbations***. On this basis, it is possible to optimize the robustness of a system, once given its scope and environmental conditions.

To this purpose, it should be considered that while trade-offs between robustness and fragility can always be optimized for a system designed for a precise scope, the same is not possible for living organisms, missing in this case the possibility to define what we mean for optimization [2,12]. In fact, in the same environment many different life forms can easily coexist, all optimized for a niche, from bacteria to plants to large predators, each one with its own robustness and fragility, each one optimized from evolution and time in a different form.

We speak of *noise* in presence of ***stochastic fluctuations in the quantitative parameters that rule the functioning of living systems at diverse level***. In mathematical terms, a measure of noise can be estimated by means of the Fano factor that shows the deviation of a stochastic system from a Poissonian behavior [3]. Even if noise can be often perceived as undesirable as bringing disturbing and deviant effects, it is clearer now how stochastic mechanisms are ubiquitous and inherent in biological systems: genetically identical cells can show very diverse behavior and distant phenotypic choices, due to the fact that gene regulatory networks are intrinsically noisy. Recent studies [3,4,14,15,16,17] have revealed that single gene/protein fluctuation is mainly determined at the translational level. The same concentration of a protein can be obtained in two ways: 1) with low transcription-high translation rates, few mRNA are produced, each one in turn producing a large and variable burst of protein production. This way results to be very noisy, with large fluctuations in protein production. The second way consists in high transcription-low translation rates: in this case, many mRNA are produced, each of them producing few proteins copies. This second procedure is energetically inefficient but the protein stream is much more steady and controlled, with small fluctuations. It should be noted that this translation inefficiency can be sustained with the beneficial reduction of noise. Thus, translation efficiency seem to be the predominant source of phenotypic noise. The same studies [4] have revealed that proteins that respond to environmental changes are "noisy", while those involved in protein synthesis are "quiet". Sometimes, intrinsic noise of a regulator can increase the sensitivity of signal transmission [3]. In parallel, the importance of noise should be considered and better studied in the case of the complex defense mechanisms of immune response. Probably, noise can play an important role in increasing the immune system response in both positive (reaction to pathogens) and negative cases (autoimmune diseases).

Another example of pervasive architecture in biology seem to be represented by the so called "***bow-tie***" structure. There are many cases in biology as well as in engineering and technology [5], of this organizational principle, consisting in ***a large "fan in" (many inputs), a relatively small "node" of control and elaboration processes, and a large "fan out" of products***. Bacterial metabolic networks [5, 18] clearly represent such structure, with many nutrients catabolized in few carriers (ATP, NADH, etc.) and precursors, in turn synthesized in a larger quantity of "building blocks" (amino acids, fatty acids, etc.). Their modularity and shared controls make these networks robust and reliable, yet with their fragility, because the same robust structures can be "abused" from pathological processes to spread and amplify through the network (for example, tumors just upregulate normal physiological processes) [5]. Inside the general bow-tie architecture, the concepts of standard protocols (for example in signal transmission), modularity (reusable blocks) and general-purpose machinery (adaptable to perform different tasks even if with different efficiency, thus recalling degeneracy) makes systems robust and together evolvable, and thus universally successful in many fields, from biology to technology to economy.

Besides these concepts, the formalization of a sort of "***grand unification theory***" in systems biology remains a challenging task that will require a consistent effort, first in identifying the fundamental principles and architecture of biological systems functioning, then in the integration of models, computational methodologies and experimental data [19], and finally in the formulation of a valuable unifying perspective, conceptually as well as mathematically. Thus the identification and the study of general architectures, standards, protocols and basic structures remains a necessity for the advancement of knowledge of immune system and biological systems functioning.